# Field-Free Spin-Orbit Torque driven Switching of Perpendicular Magnetic Tunnel Junction through Bending Current

*Vaishnavi Kateel[1,2]\*, Viola Krizakova[3], Siddharth Rao[1]\*, Kaiming Cai[1], Mohit Gupta[1], Maxwel Gama Monteiro[1,2], Farrukh Yasin[1], Bart Sorée[1,2], Johan De Boeck[1,2], Sebastien Couet[1], Pietro Gambardella[3], Gouri Sankar Kar[1], Kevin Garello[1,4]\**

1. IMEC, 3001 Leuven, Belgium; 2. KU Leuven, 3001 Leuven, Belgium; 3. Department of materials, ETH Zurich, 8093 Zürich, Switzerland; 4. Univ. Grenoble Alpes, CEA, CNRS, Grenoble INP, SPINTEC, 38000 Grenoble, France

**ABSTRACT:** Current-induced spin-orbit torques (SOTs) enable fast and efficient manipulation of the magnetic state of magnetic tunnel junctions (MTJs), making it attractive for memory, in-memory computing, and logic applications. However, the requirement of the external magnetic field to achieve deterministic switching in perpendicular magnetized SOT-MTJs limits its implementation for practical applications. Here, we introduce a field-free switching (FFS) solution for the SOT-MTJ device by shaping the SOT channel to create a "bend" in the SOT current. The resulting bend in the charge current creates a spatially non-uniform spin current, which translates into inhomogeneous SOT on an adjacent magnetic free layer enabling deterministic switching. We demonstrate FFS experimentally on scaled SOT-MTJs at nanosecond time scales. This proposed scheme is scalable, material-agnostic, and readily compatible with wafer-scale manufacturing, thus creating a pathway for developing purely current-driven SOT systems.

Electrical manipulation of magnetic tunnel junctions (MTJs) is gaining importance for embedded memory applications and spin-based logic due to the non-volatile nature of ferromagnets, CMOS compatibility, and high operational speed[1]. In particular, MTJs switched with spin-orbit torque (SOT) is promising owing to its high endurance ($> 10^{14}$), large energy efficiency (< 1 fJ), and fast switching speed (< 1 ns) characteristics[2–7]. In SOT-MTJs, the spin Hall effect in the SOT channel (heavy metal) and the interfacial Rashba interaction at the SOT channel/ferromagnetic free layer (FL) interface creates a spin current with in-plane spin polarization, which induces an instantaneous torque on the magnetization. To enable ultra-fast SOT-induced switching, perpendicularly magnetized MTJs (pMTJs) are preferred[4]. However, in such a configuration, the generated torques cannot intrinsically differentiate between the two states of pMTJs, thus resulting in non-deterministic switching[8]. To enable SOT-induced deterministic switching, the time-reversal symmetry needs to be broken; this is traditionally achieved with an external in-plane magnetic field ($B_{ip}$) along the electric current direction. However, using this external field hinders the potential of SOT-MTJs for embedded memory applications.

Various alternatives have been proposed in the literature to remove or substitute this external field, which can be broadly grouped into four categories[9]: (i). Built-in magnetic fields by exchange bias[10] or a decoupled in-plane magnet grown in the pMTJs structure[11–13]; (ii). Additional spin currents are produced with crystalline materials (PtCu[14,15], WTe$_2$[16–18]), Rashba field modulation[19,20], or competing spin Hall angle[21] (iii). Hybrid approaches combine domain wall motion[22], spin-transfer torque[23], Dzyaloshinskii-Moriya interaction (DMI) modulation[24], the spin current generated from an in-plane magnetized ferromagnetic layer[25] or coupled nanomagnets[26] with SOT-induced switching, and finally (iv). Structural asymmetry is created by a gradient of the FL thickness [27,28] or the SOT channel thickness [29], tilting the easy axis of the FL[30] and engineering the FL geometry[31].

In this study, we propose a new field-free switching (FFS) scheme for SOT-pMTJs by introducing a geometric curvature in the SOT channel (Figure 1a). This curvature creates a structural asymmetry in the system with the flow of inhomogeneous charge current in the SOT channel, which modifies the flow and polarization of spin current. These modifications in spin current further cause an inhomogeneous SOT on the FL, thus breaking the time-reversal symmetry. We demonstrate deterministic FFS for current pulses down to 5 ns and elucidate the underlying mechanism with micromagnetic and finite-element simulations. Importantly, this approach has no additional restriction on the material choices for the SOT channel or the FL. Manipulating magnets by inducing a curvature in the SOT channel holds exciting potential for practical application due to its scalability, integration friendliness, and reproducibility.

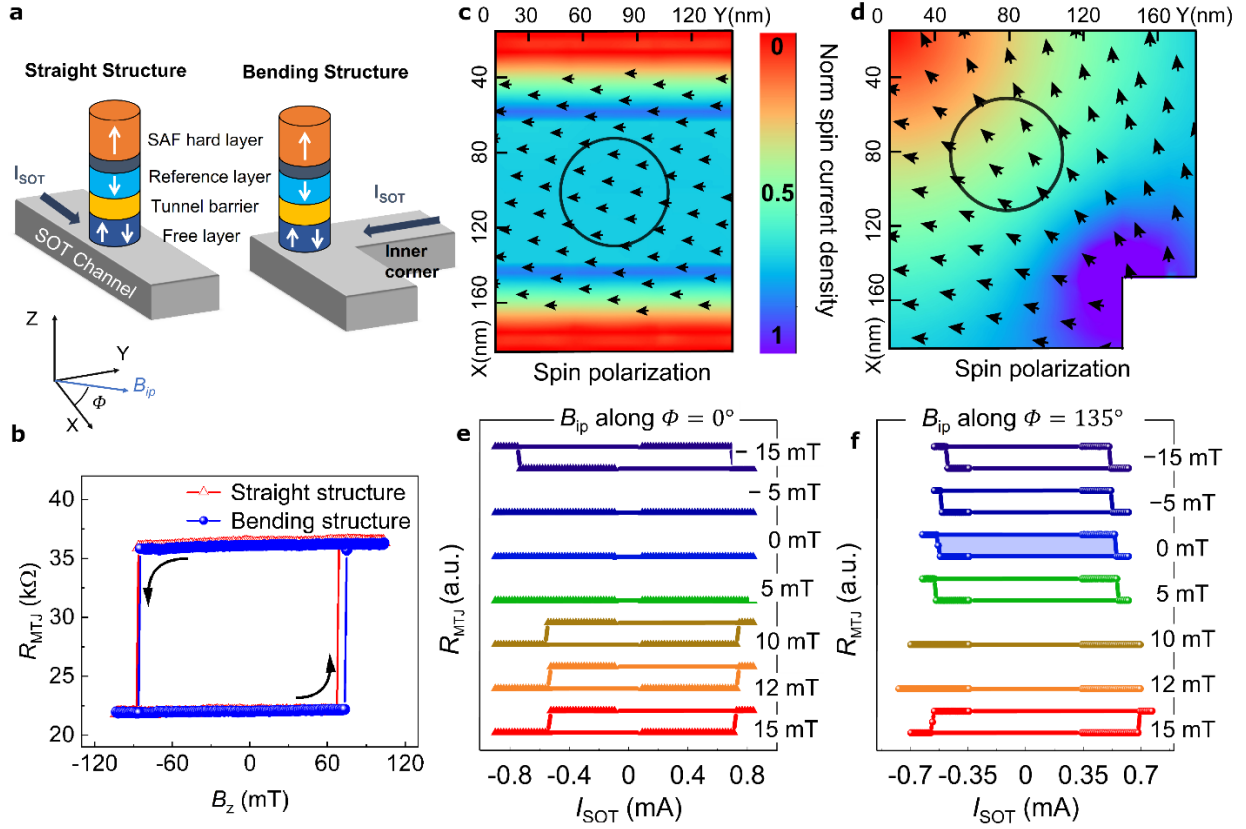

*Figure 1. FFS in bending structure and comparison with straight structure.* **(a)** Schematic showing placement of pMTJ on the SOT channel for straight and bending structures. **(b)** Hysteresis loop of the free layer in straight and bending structures. **(c and d)** Simulated color map of the normalized spin current density ($J_S$) in the SOT channel with the arrows indicating the direction of the spin polarization ($\sigma$); (*c*) The straight structure shows $\sigma$ pointing along y, and a uniform $J_S$ around the MTJ with variations near the edges due to the proximity to VIA contacts. (*d*) The MTJ in the bending structure experiences a $J_S$ gradient and change in the direction of $\sigma$ due to current bending at the corners. *(e and f)* Average of 20 DC switching loops for the varied magnitude of $B_{ip}$ along (*e*) $\Phi = 0°$ for straight structure and (*f*) $\Phi = 135°$ for bending structure.

---

In this work, we performed the experiments on our baseline top-pinned SOT-MTJs fabricated on 300 mm wafers and annealed to 300 °C, as reported in ref.[6]. The MTJ consists of a 0.9 nm CoFeB free layer, a MgO barrier with a resistance area product of ~51 Ω·µm$^2$, and a 1.1 nm CoFeB reference layer, which is pinned to a synthetic antiferromagnetic (SAF) hard layer. The hard layer pins the reference layer in the −z direction such that the MTJ is in a low resistive parallel (P) or high resistive antiparallel (AP) state when the FL magnetization points toward the −z (+z) direction, respectively. MTJ pillars have a circular cross-section with nominal diameters of 60, 80, and 100 nm. The MTJs are deposited on a 3.5 nm thick β-W SOT channel with a resistivity of 160 μΩ cm and an effective spin Hall angle ($\theta_{SH}$) estimated to be −0.43 from our second Harmonic measurement[32]. All properties for each dimension are summarized in Supplementary Information I. Henceforth, in this article, we focus on the results from 60-nm MTJs.

The device designs used in our study (Figure 1a) are standard straight-line SOT channel with MTJ positioned at the center of the channel, referred to as a straight structure (as reference structures), and a cornered SOT channel with a MTJ positioned on the corner, referred to as bending structure. Both structures have a SOT channel width of 140 nm; meanwhile, the length in the bending structure is longer than in the straight structure. This longer length in bending structures causes higher resistance (4350 Ω) than straight structures (500 Ω). The MTJ shows a square hysteresis loop (Figure 1b) with similar magnetic properties for both structures. The median values for over 20 structures of each type are a tunnel magnetoresistance ratio (TMR) of ~73 ± 4 %, a coercive field ($B_c$) of ~ 60 ± 6 mT, an offset field along the z direction $|B_{off}|$ ~ 6.8 ± 4 mT (favoring AP state), and an anisotropy field $B_k$ ~180 ± 3.4 mT.

The surface profile of the magnitude of the spin current ($J_S$) along z and the direction of spin current polarization (σ) for both structures are simulated using COMSOL (Supplementary Information II). For straight structures, when a charge current is applied along +x, due to the spin Hall effect, a constant $J_S$ and a homogenous distribution of σ in the −y direction is observed (Figure 1c). In the bending structures, the reorientation of electron flow at the inner corner causes a rotation of σ(**r**), which modifies the SOT direction across the FL[33]. Additionally, the spatial distribution of charge current ($J_C(\mathbf{r})$) results in a nonuniform amplitude of $J_S$ (Figure 1d), as $J_S(\mathbf{r}) \sim \frac{\hbar}{2e}(J_C(\mathbf{r}) \times \boldsymbol{\sigma}(\mathbf{r}))\theta_{SH}$, which results in gradient of $J_S(\mathbf{r})$ in the *xy* plane with highest amplitude of $J_s(\mathbf{r})$ at the inner corner. This generated spin current gradient is independent of the direction of charge current due to higher current accumulation at the inner corner of SOT channel. Importantly, due to the inhomogeneous $\boldsymbol{\sigma}(\mathbf{r})$ and $J_S(\mathbf{r})$, a spatially varying SOT is experienced by the FL; thus, it breaks the time-reversal symmetry, and it allows deterministic switching.

We perform SOT-induced magnetization switching by passing direct-current (DC) of variable amplitude ($I_{SOT}$) through the SOT channel. A post-pulse readout of the FL state is sensed by applying a low read current ($I_{read} \sim 5$ µA) through the MTJ. During the switching, $\boldsymbol{B}_{ip}$ can be applied in the *xy* plane with a maximum amplitude of 40 mT. Figure 1e shows typical $R_{MTJ}$-$I_{SOT}$ switching loops for straight structures obtained at different $\boldsymbol{B}_{ip}$ applied along the *x*-direction. The switching conditions are as expected for a negative spin Hall angle; at zero external fields, the device does not switch deterministically, and a minimum $\boldsymbol{B}_{ip} \sim \pm 10$ mT is required to define the switching polarity. When a positive $I_{SOT}$ is applied, the MTJ state switches from anti-parallel (AP) to parallel (P) state for $\boldsymbol{B}_{ip} < -10$ mT, whereas P-AP transition for $\boldsymbol{B}_{ip} > 10$ mT. Thus, SOT switching for straight structures with positive or negative $\boldsymbol{B}_{ip}$ generates clockwise (CW) or counterclockwise (CCW) switching loops, respectively. The average switching current ($I_{SW}$) over our devices is ~500 µA, corresponding to a switching current density ($J_{SW}$) of $9.5 \times 10^7$ A cm$^{-2}$ for $\boldsymbol{B}_{ip} = -10$ mT.

Remarkably, for the bending structures shown in Figure 1f, deterministic bipolar switching occurs at $\boldsymbol{B}_{ip} = 0$ mT, thus confirming the proposed concept of time-reversal symmetry breaking due to inhomogeneous SOT below the FL and $\boldsymbol{J}_S$ gradient. FFS loops in bending structures have CW polarity, the same as the straight structures with $\boldsymbol{B}_{ip} < 0$, indicating the direction of the effective SOT field ($\boldsymbol{B}_{eff}$) created by bending the SOT channel. When applying $\boldsymbol{B}_{ip} \sim 10$ mT (at

−45° in *xy* plane, which is the effective $J_C$ direction), the device shows stochastic switching, whereas a further increase of $B_{ip}$ causes the switching polarity to reverse from CW to CCW loops. This change in switching polarity suggests that for $B_{ip} > 15$ mT, $B_{ip}$ dominates over $B_{eff}$. The average $I_{SW}$ of the bending structure is like straight structures, with ~530 µA, corresponding to $J_{SW} = 1 \times 10^8$ A cm$^{-2}$ at $B_{ip} = 0$ mT. We observe FFS using MTJ sizes of 80 and 100 nm on a W-based SOT channel. Importantly, we observe FFS with Pt-based MTJs, but with opposite switching polarities. Since Pt has an opposite spin Hall angle to W, this validates that SOT physics dominates the FFS in bending structures (Supplementary Information III).

To investigate the direction and the magnitude of $B_{eff}$ in the bending structure, we perform switching measurements with an angular variation of the $B_{ip}$ and compare it to straight structures. We record average switching loops in the presence of a $|B_{ip}| = 25$ mT with varying directions in the *xy* plane from $-180 \leq \Phi \leq 180°$. We observe a cosine angular dependence of $I_{SW}$ with $\Phi$, which is due to the variation of angle between $B_{ip}$ and the antidamping SOT field ($B_{AD}$)[8,34]. The opposite trend in variation of $I_{SW}$ for P-AP and AP-P transition relates to the interplay between the field-like field ($B_{FL}$) and $B_{ip}$, which add or subtract from each other[35]. The direction of $B_{eff}$ can be identified by the angular variation of $B_{ip}$ since the interplay between $B_{ip}$, $B_{AD}$, and $B_{FL}$ determines the region for the lowest switching current and stochastic (random) switching[36]. In the straight structure (Figure 2a), the region for stochastic switching occurs at ~ ±90°, and the least switching current occurs at 180° for CW (0° for CCW switching loop). The $B_{eff}$ is determined by the sum of $B_{FL}$[34] or $B_{AD}$[35] in SOT-induced switching in the presence of $B_{ip}$ and is along the direction of current (180° for CW switching loops).

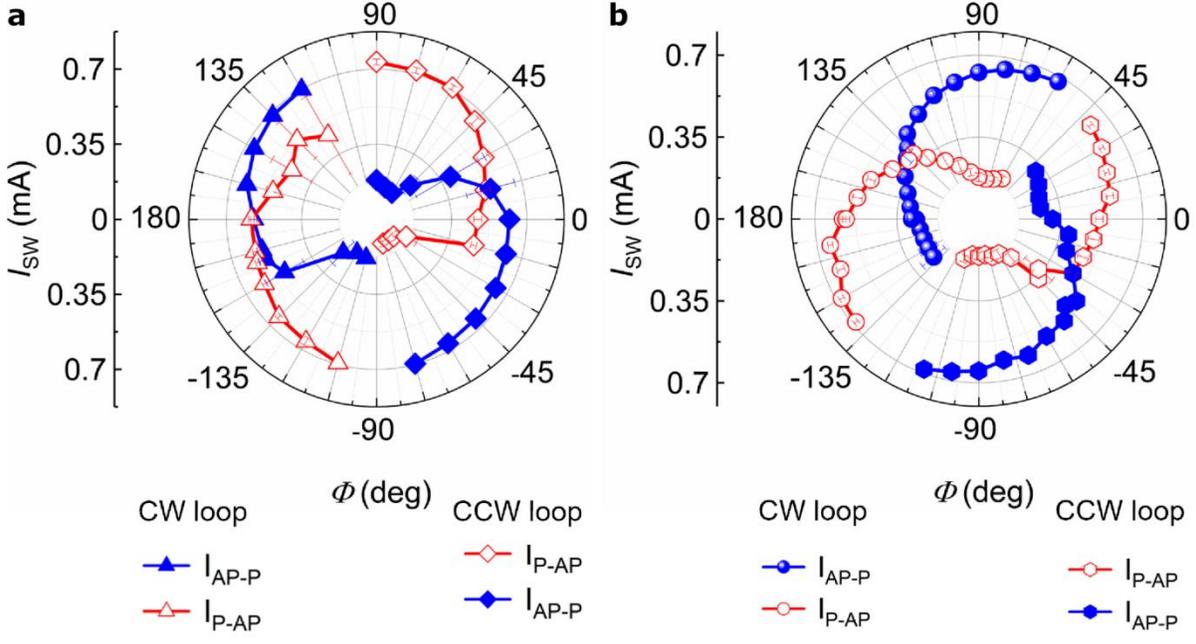

*Figure 2. Critical switching currents as a function of angular variation of $B_{ip}$ in xy plane with $|B_{ip}|$ = 25 mT. (a) Polar plot for straight structures shows CW (CCW) switching loop with the least switching current for both transitions at $\Phi = 180°$ (or −15°), and there is no switching at −90° (or 110°). (b) Polar plot for bending structures shows CCW (CW) switching loops for angles −120° ≤ $\Phi$ ≤ 40° ( $\Phi$ ≥ 60°, $\Phi$ ≤−150°). The least switching current and stochastic switching are at $\Phi = 135°$ (−30°) and 45° (−130°), respectively.*

For the bending structure (Figure 2b), the region showing the CW switching loop is for angles between −140° and 50°. The minimum switching current required for both transitions are at 135° and −30° for CW and CCW switching loops, respectively. Using the same analogy for bending structures, we can conclude that the $\boldsymbol{B_{eff}}$ contributing to FFS will be along 135°. We have used this understanding for measuring the average $R_{MTJ}$-$I_{SOT}$ switching loop in Figure 1e, f.

To gain insight into the underlying switching mechanism, we performed micromagnetic simulations for the bending and the straight structures. In the micromagnetic framework, we consider the exchange interaction, anisotropy, damping parameter, generated SOT (both antidamping SOT and field-like SOT), Oersted field, magnetostatic interaction, and DMI. We first perform a finite-element simulation to calculate the spatial distribution of $J_C$ induced by the bent SOT channel, resulting in the gradient in $J_S$, inhomogeneous $\sigma$ (Figure 1d), and the Oersted field. The magnetization trajectory in our simulation is recorded by initializing the FL along +z (or −z) and followed by a relaxation of 0.1 ns. Then, the $I_{SOT}$ switching pulse is applied to 1 ns (indicated by the yellow region in Figure 3 a, b), and the system is allowed to relax for another 6 ns with the current pulse turned off. For both straight structure (Figure 3a) and bending structure (Figure 3b), deterministic switching occurs with AP-P transition for $I_{SOT} > 0$ (inset graph showing P-AP transition for $I_{SOT} < 0$), in the presence of $B_{ip} = -20$ mT and absence of field, respectively. The FFS is reliable and shows fixed switching polarity confirming the CW loops observed in the experiments. Notably, there is no switching when the FL is forced to have opposite switched polarity (setting initial P (AP) state for $I_{SOT} > 0$ ($I_{SOT} < 0$)).

In Figure 3c, the snapshots of magnetic configuration at different times for straight structures show the reversal by chiral domain nucleation at the edge followed by domain wall propagation across the FL, typical to SOT-MTJs of diameter ≥ 50 nm with perpendicular magnetic anisotropy (PMA)[7,37]. The nucleation point is different for both transitions as it is mainly driven by DMI[35] and SOT fields, and it has a four-fold symmetry that depends on the interplay of the in-plane field and current polarities. Interestingly, nucleation occurs on the spot closer to the inner corner in bending structures (Figure 3d) regardless of initial magnetization or current direction in the SOT channel. Given the spatial current distribution, nucleation is always favored at the inner corner of FL due to the higher spin current density for both polarities, and it is followed by a domain wall propagation until the current pulse is turned off. The domain wall propagation is attributed to an inhomogeneous SOT[31] due to the continuous spin polarization direction change. Once the pulse is turned off, with the support of exchange interaction, anisotropy, and DMI, the magnetization relaxes to a uniform ground state determined by the initial nucleated domain orientation. We further investigated the roles of spin current gradient, inhomogeneous spin polarization, and DMI on the switching of bending structure (Supplementary Information IV.i). Notably, the spin current gradient contributes to fixing the nucleation point at the inner corner of the SOT channel for both current polarities. The domain wall dynamics are dictated by the direction of spin polarization and the strength and direction DMI. The domain wall magnetization experiences an inhomogeneous SOT due to spatially varying spin polarization; its magnetization orientation depends on the direction of current and initial magnetization of the FL, thus enabling the differentiation between the two transitions. The DMI strength and chirality impacts the post-nucleation dynamics with only positive DMI (promoting right-handed domain wall chirality), enabling switching for our current bending structure design. The switching polarity is determined by the combination of DMI chirality and SOT induced on the domain wall magnetization due to spatially varying spin polarization. The fixed nucleation point determined by the spin current gradient and followed by complete switching can also be observed for different pulse parameters, MTJ sizes and MTJ locations on the SOT channel, indicating reliable operations (Details discussed in Supplementary Information IV.ii). We also simulate the reversal dynamics at an elevated temperature and observe no impact on the nucleation site or the switching trajectory compared to the 0 K simulations, indicating that the switching is reliable against random thermal fluctuation. This corroborates our assumption that the deterministic nature of the FFS process is due to the spatially varying SOT assisted by DMI.

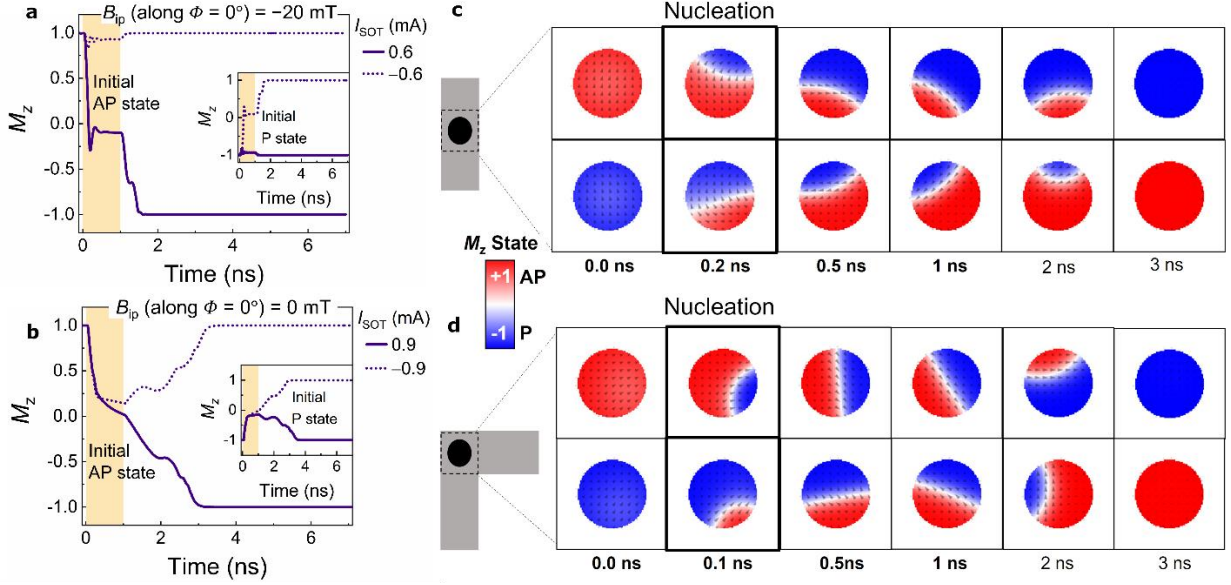

*Figure 3. Micromagnetic simulation of magnetization switching. (a and b)* Normalized $M_Z$ trace showing switching from AP state for positive current (no switching for negative current) and inset showing P state switching for negative current (no switching for positive current) in *(a)* straight structure in presence of $B_{ip} = -20$ mT *(b)* bending structure in the absence of $B_{ip}$. *(c and d)* Still frame of magnetization for both transitions at various times during switching indicating *(c)* nucleation point changing with the direction of the current for straight structures in the presence of $B_{ip} = -20$ mT *(d)* FFS with fixed nucleation point at the inner corner determined by the spin current gradient for both the transitions.

In addition, we investigated the role of inhomogeneous $B_{FL}$ and the Oersted field generated due to the curvature of the SOT channel by performing simulation in the presence and absence of these parameters, there is little change between the magnetization trajectory (Supplementary Information IV.iii). The damping parameter is varied in magnitude; there is no significant variation in the switching trace recorded in the simulations. Meanwhile, the spatial variation of the magnetic parameters due to inhomogeneous Joule heating of the SOT channel has a moderate influence; we observe that it only lowers the critical switching current and does not impact the switching dynamics for the bending structure. The magnetization reversal in bending structure is mainly influenced by inhomogeneous current distribution below the SOT and chirality of the DMI;

other factors such as the Oersted field, $B_{FL}$, damping parameter, and inhomogeneous Joule heating of the SOT channel play no role in time-reversal symmetry breaking but only contribute towards enhancing (reducing) the switching reliability.

Notably, we experimentally compare the pulsed switching of bending structure at $B_{ip}$ = 0 mT and the switching of straight structure at the various in-plane field in Figure 4a (More details on experimental setup are given in Supplementary Information III). The pulsed switching confirms that the proposed FFS mechanism seems not limited by switching speed as we can deterministically switch at 5 ns with $J_{SW}$ = 1.5 × 10$^8$ A cm$^{-2}$. Unfortunately, for pulse width ($\tau_p$) shorter than 5 ns, switching is limited by the large SOT channel resistance of the bending structures and our experimental setup voltage delivery. At longer $\tau_p$ (> 50 ns), the $I_{SW}$ for bending structure ($B_{ip}$ = 0 mT) is like straight structure ($B_{ip}$ = −12.8 mT along $x$-direction) as the stochastic thermal fluctuations contribute to lowering the required $B_{eff}$ ($\propto J_S$) for bending structures. The difference in switching current increases at shorter pulses between both structures as the $B_{eff}$ in bending structures is current amplitude-dependent; in contrast, $B_{eff}$ is static ($B_{eff} = B_{ip}$) for straight structures. We also note that reliable switching at 10 ns is possible for bending structures with larger MTJ diameters of 80 nm and 100 nm (Supplementary Information III).

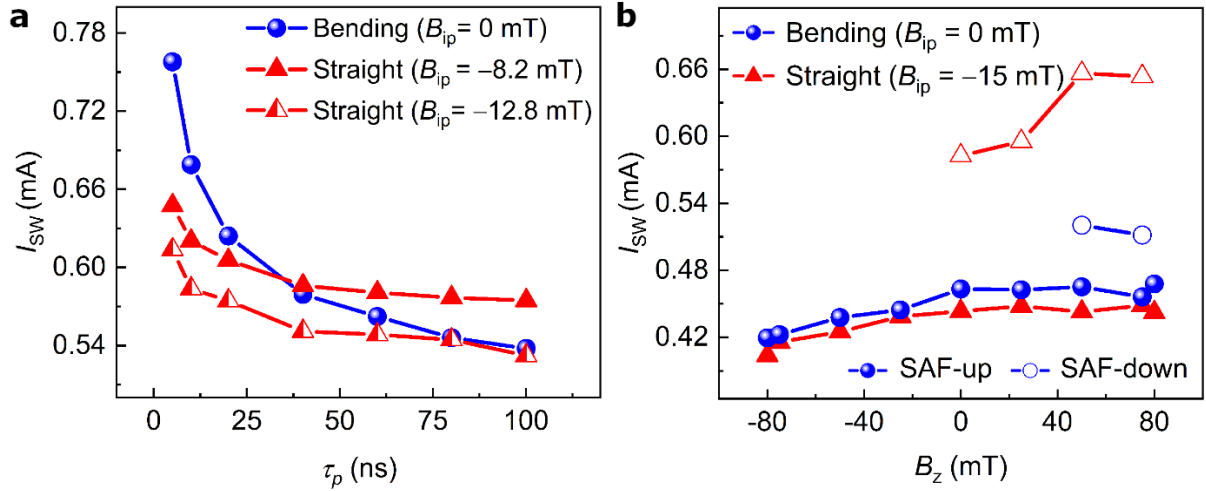

*Figure 4. Switching performance of bending structures. (a)* $I_{SW}$ *as a function of the pulse width* ($\tau_p$) *in bending structures* ($B_{ip}$ = 0 mT), *straight structures* ($B_{ip}$ of −8.2 mT and −12.8 mT along $\Phi$ = 0°). *(b)* $I_{SW}$ *(direct current) as function* $B_z$ *for bending structures at* $B_{ip}$ = 0 mT *and straight structures at* $B_{ip}$ = −15 mT *for hard layer orientation* +$z$ *(SAF up) and* −$z$ *(SAF down).*

Given the switching dynamics observed in the simulations, and the importance of a z field on DW displacement[38], we measured the switching as a function of the direction of the stray field produced by the hard layer and the out-of-plane field ($B_z$), as shown in Figure 4.b. Bending structures devices with hard layer magnetized in +$z$ (SAF up) shows reliable switching for a wide range of $z$-field whereas SAF down structures show switching only for large positive z-field. This shows that the stray field generated from the hard layer impacts the switching dynamics by either assisting or competing with $B_{AD}$ on the domain wall magnetization during the reversal[38] (More details in Supplementary Information V). This effect is more pronounced in bending structures than straight structures as the magnetization of the domain wall in the bending structure is determined by the DMI (intrinsic property), whereas for the straight structure, it is controlled by the magnitude and direction of $B_{ip}$.

**DISCUSSION AND CONCLUSION**

Finally, to access impact of bending structures on the architecture design, we performed design-to-technology co-optimization (DTCO) analysis for bending structures and compared it to other embedded memory technologies such as static-RAM, STT-MRAM, and SOT-MRAM at the 5 nm technology node[39] (Supplementary Information VI). Notably, we observed that bending the SOT channel does not impact the area per bit compared to the standard SOT-MRAM, with the added advantage of field-free switching and no penalty on the switching current.

However, various aspects need to be investigated to improve the switching performance of bending structures in terms of design and material characteristics for making it a promising FFS solution. Firstly, understanding the impact of the stray field generated from the SAF hard layer on FFS and engineering the hard layer to assist in the domain wall dynamics is required. Secondly, optimizing interface properties between the FL and SOT channel, such as DMI and $\theta_{SH}$, will cause an increase in spin current gradient and improve the domain wall dynamics. Thirdly, redesigning the structure by shortening the SOT channel will reduce the resistance, limit the breakdown of the device, and give the opportunity to study a shorter pulse regime. Lastly, placing the MTJ closer to the inner corner will cause a larger gradient in the current below the MTJ. The expected reduction in the critical current is ~25%, if MTJ is closer to the inner corner of the SOT channel by 18 nm compared to our current design. The DTCO analysis for this optimized location of MTJ shows similar properties as the straight structure.

To conclude, we proposed and demonstrated a new FFS concept by introducing a curvature in the SOT channel in a SOT-pMTJ with no impact on the static properties of the MTJ stack. Deterministic field-free switching is achieved down to nanosecond pulses due to time reversal symmetry breaking created by inhomogeneous SOT below the free layer. The switching mechanism can be explained by the creation of a fixed point for nucleation determined by the geometry of the SOT channel, followed by a complex domain wall propagation supported by inhomogeneous SOT. Bending structure design opens a new route for realizing FFS in SOT-pMTJs, it is scalable, material agnostic, and suitable for large-scale integration.

**SUPPLEMENTARY INFORMATION:**

- I. Magnetic properties of bending structure
- II. Finite element method simulation
- III. SOT induced switching in bending structures
- i. Experimental setup for DC SOT switching
- ii. Experimental setup for pulsed SOT switching
- iii. Field-free switching
- IV. Micromagnetic Simulation:
- i. Switching mechanism of bending structures
- ii. Switching robustness of bending structures
- iii. Change in simulation parameters.
- V. Influence of SAF Hard layer
- VI. Bending structure design analysis

**METHODS**

Sample fabrication

The devices were fabricated on IMEC's 300 mm MRAM pilot line. SOT-MTJ stacks are deposited on pre-processed wafers made of W-VIA bottom electrodes embedded in $SiO_2$. The SOT-MTJs were deposited in-situ at room temperature by physical vapor deposition in a 300 mm cluster EC7800 Canon-Anelva tool and subsequently annealed for 30 minutes at 300 °C in the presence of the perpendicular magnetic field of 1 T. The material composition of top pinned MTJ stack for W/CoFeB is W(3.5)/Co$_{20}$Fe$_{60}$B$_{20}$(0.9)/MgO (~1)/Co$_{17.5}$Fe$_{52.5}$B$_{30}$(1.1)/W(0.3)/Co(1.2)/Ru(.85)/Co(0.6)/Pt(0.8)[Co(0.3)/Pt(0.8)]$_6$Ru(5) (units in nm). Then, the MTJ pillar was patterned using 193 nm immersion lithography followed by ion-beam etching at normal incidence with etch stop condition optimized. The SOT channel is intact without producing a side wall short across the barrier. The copper electrodes were fabricated using a dual damascene process to make the top and bottom electrical connections.

**Author Information**

**Corresponding Author:**

Vaishnavi Kateel- IMEC, 3001 Leuven, Belgium, KU Leuven, 3001 Leuven, Belgium; Email: vaishnavi.kateel@imec.be

Siddharth Rao - IMEC, 3001 Leuven; Email: siddharth.rao@imec.be

Kevin Garello- IMEC, 3001 Leuven, Univ. Grenoble Alpes, CEA, CNRS, Grenoble INP, SPINTEC, 38000 Grenoble, France; Email: kevin.garello@cea.fr

**Acknowledgment**

This work is supported by imec's Industrial Affiliation Program on MRAM devices, ECSEL Joint Undertaking Program (grant No. 876925– project ANDANTE), and Swiss National Science Foundation (Grants No. 200020_200465).

**Contribution**

K.G. planned and supervised the project. Va.Ka. Performed the measurement on devices designed by K.G.. F.Y. was in charge of the device integration process. The experimental data were analyzed by Va.Ka., S.R., K.C., and K.G.. Micromagnetic simulations were performed by Vi.Kr., and M.G.M. M.G. performed the design-to-technology co-optimization analysis. Va.Ka., S.R., and K.G. wrote the manuscript. All the authors discussed the data and commented on the manuscript.

**Competing interests**

**Data Availability**

The data that support the plots within this paper and other finding of this study are available from the corresponding author upon reasonable request.